# Towards searching for Majorana fermions in topological insulator nanowires

Hong-Seok Kim, and Yong-Joo Doh[*]

*Department of Physics and Photon Science, Gwangju Institute of Science and Technology, Gwangju 61005, Korea*

## Abstract

Developing a gate-tunable, scalable, and topologically-protectable supercurrent qubit and integrating it into a quantum circuit are crucial for applications in the fields of quantum information technology and topological phenomena. Here we propose that the nano-hybrid supercurrent transistors, a superconducting quantum analogue of a transistor, made of topological insulator nanowire would be a promising platform for unprecedented control of both the supercurrent magnitude and the current-phase relation by applying a voltage on a gate electrode. We believe that our experimental design will help probing Majorana state in topological insulator nanowire and establishing a solid-state platform for topological supercurrent qubit.

*Keywords*: Majorana fermion, topological insulator nanowire, Shapiro steps

# 1. INTRODUCTION

A Josephson junction (JJ) can be made by sandwiching a thin layer of a nonsuperconducting material between two superconducting electrodes [1]. Here, supercurrent can flow through the intervening layer without any dissipation via the superconducting proximity effect. Dynamics of a JJ, i.e. current-voltage characteristics, can be understood by the one of a fictitious phase particle confined in a washboard-type potential well. The phase particle is expected to have discrete energy levels inside the potential well due to a quantum confinement effect, which is like an atom or quantum dot (so-called an artificial atom). Thus we can call a JJ as a superconducting artificial atom [2]. Macroscopic quantum tunneling (MQT) which is observed in a JJ provides a direct evidence to prove its quantum nature. The energy-level spacing inside the Josephson potential well is determined by the Josephson coupling energy, $E_J = \hbar I_c/2e$, where $\hbar$ is Planck's constant, $I_c$ a maximum value of supercurrent, and $e$ an electric charge. Since a linear superposition of the basis states in a two-state quantum system constitutes a qubit (or quantum bit), a quantum mechanical analogue of the classical bit, a single JJ provides us a unit of quantum information to be useful for quantum computing [3].

There are three prototypical superconducting qubits, which are the phase, charge and flux qubits. The two basis states of |0> and |1> are mapped to the different quantum numbers of each qubit, respectively. In the phase qubit, discrete energy levels in a washboard potential are used as the basis states to form a qubit, as explained above. In the charge qubit, discrete quantum states are assigned by an integer number of Cooper pairs on a superconducting island (or a Cooper pair box). In the flux qubit, the two lower states correspond to a clockwise and a counterclockwise current states in a supercurrent loop (or a SQUID loop). In addition, an integrated quantum circuit of the superconducting qubit and microwave cavity forms a transmon qubit [4]. Recently the maximum number of qubits integrated in a chip is nine [5] and the coherence time ($T_1$) reaches to ~ 40 μs in a planar on-chip qubit [6].

We note that all the above superconducting devices are conventional tunneling-type junctions made of Al/Al$_2$O$_3$/Al. Then there occurs inevitable drawbacks in the superconducting junctions. Firstly, the stable operation of the superconducting qubits is limited at very low temperatures ($\leq$ 10 mK), which is far below the superconducting transition temperature of Al (~ 1 K). Secondly, controlling E$_J$ is only possible by applying an external magnetic field to modulate I$_c$, which is detrimental to the scalability of qubits. Thus it is required to develop a new device scheme, which can be easily integrated into a quantum circuit and also work at higher temperatures with maintaining quantum coherence. Here we propose an idea to utilize topological insulator nanowires as a platform to develop novel quantum information devices, which would enable topological quantum computation.

## 2. BACKGROUND

Semiconductor-based field-effect transistors (FETs) allow us to control normal current flow through the conducting channel by the application of gate voltage. In principle, it is also possible to control the supercurrent flow through a semiconductor using the gate voltage, which is called as a Josephson field-effect transistor (or supercurrent transistor). In the laboratory, however, it is very hard to observe the supercurrent flow through a semiconductor, due to a formation of a Schottky barrier at the interface between a metal and a semiconductor. In 2005, we have successfully developed a new device scheme of nano-hybrid supercurrent transistor using InAs semiconductor nanowire [7]. This work has been applied to other semiconductor nanostructures, such as carbon nanotube [8], graphene [9], Ge/Si nanowire [10], graphene p-n device [11], and topological insulators (TIs) [12]. These semiconductor-based supercurrent transistors will pave a way to develop various quantum information devices, such as a gate-tunable superconducting quantum interference device (SQUID) [13]. In particular, Majorana state has been already probed using InAs [14] and InSb [15] semiconductor nanowires contacted with conventional *s*-wave superconductors, which would be a corner stone for topological quantum information process. Furthermore, gate-tunable superconducting qubits have been also developed using InAs nanowires, opening a new research field of nano-hybrid superconducting qubits, so-called gatemon [16, 17].

TIs are bulk insulators including metallic (and topological) surface states, which are topologically protected by time-reversal symmetry [18]. The surface states are also known to be spin-helical, meaning that the electron spin is aligned parallel to the surface and normal to the momentum. Since the spin orientation of the surface electrons are locked perpendicular to their translational momentum ("spin-momentum helical locking"), the metallic surface states in TIs are protected from backscattering by chiral spin texture and thus expected to exhibit highly quantum-coherent charge and spin transport, making this TI material to be promising for novel quantum information devices. It has been theoretically expected that TIs combined with *s*-wave superconductor can provide useful platforms for creating and manipulating Majorana states [19], which are essential for topological quantum computation.

Although spin-resolved ARPES (angle-resolved photo-emission) studies has successfully verified the spin-helical state in TIs, the electrical transport study failed to isolate the topological surface state (TSS) due to the residual bulk carriers created by material imperfection. This issue can be tackled by using TI nanowires, where the surface state contribution can dominate the whole electrical transport owing to the large surface-to-volume ratio. In TI nanowire experiments, the existence of TSS can be proved by measuring the periodic oscillations of electrical conductance with increasing axial magnetic field, where the oscillation period is given by a single flux quantum, $\Phi_0 = h/e$, through the cross-sectional area of TI nanowire [20]. This is caused by Aharonov-Bohm (AB) oscillations, which is an interference effect between partial waves of surface electrons encircling the magnetic flux $\Phi$. Since there are additional two terms from the formation of one-dimensional (1D) subbands along the circumference and Berry's geometric phase in TI, the dispersion relation for the TSS in TI nanowires (NWs) is as follows:

$$E(n,k,\Phi) = \pm h v_F \sqrt{\left(\frac{k}{2\pi}\right)^2 + \left(\frac{n+1/2-\Phi/\Phi_0}{L_p}\right)^2}$$

where $n$ is a quantum number, $L_p$ is the perimeter length, and $v_F$ is the Fermi velocity [21]. It should be noted that there forms a gapped state for integer flux quanta ($\Phi = 0, \pm\Phi_0, \pm 2\Phi_0, ...$) and helical linear dispersion for half-integer flux quanta ($\Phi = \pm\Phi_0/2, \pm 3\Phi_0/2, ...$), when the Fermi level is located near Dirac point. As a result, we can isolate the helical surface state and measure its electrical transport properties using TI nanowires. Forming highly transparent contacts between TI nanowire and *s*-wave superconductors will enable us to obtain gate-tunable topological supercurrent for the qubits and furthermore a unique platform for manipulating Majorana fermions [22]. We believe that the topological supercurrent qubit would pave a way for topological quantum information process based on non-Abelian exchange statistics.

## 3. EXPERIMENTAL DESIGN

Majorana fermion state in TI nanowire-based superconducting junction can be verified by several quantum electronic transport measurements such as the absence of odd-numbered Shapiro steps, $4\pi$-periodic current-phase relation and zero-bias conductance peak. A detailed research plan is forthcoming. Firstly, the JJ made of TI nanowire and *s*-wave superconductors is expected to exhibit an abnormal current-phase relation of $I_s = I_c \sin(\phi/2)$, where $\phi$ is a superconducting phase difference between two superconducting electrodes. This $4\pi$-periodic current-phase relation results in doubling of Shapiro step voltages (or absence of odd-integral Shapiro steps) under the application of an external microwave, satisfying $V_n = nhf_{mw}/e$ where $V_n$ is the *n*-th Shapiro step voltage, *n* is an integer, and $f_{mw}$ is the microwave frequency (see Fig.1a). This has been observed in a hybrid semiconductor-superconductor junctions [23], but not in a TI nanowire-based JJ yet. TI nanowires combined with conventional superconducting electrodes would provide a promising platform to measure the abnormal Shapiro steps due to Majorana states formed in TI nanowire. So far, the existence of topological surface state in TI nanowires has been verified by observing Aharonov-Bohm oscillations in various TI nanowires applied with an axial magnetic field [20, 24, 25], while the superconducting proximity effect through TI nanowires resulted in a clear observation of supercurrent [26]. We suggest Al or PbIn alloy as a superconducting electrode to form a highly transparent interface with semiconductor or TI nanowires [7, 13, 27, 28].

Secondly, the abnormal current-phase relation in TI-based JJ is expected to exhibit $4\pi$ periodicity instead of $2\pi$ one, which is observed in conventional JJ. It would be possible to measure directly such current-phase relation by forming a d.c. SQUID made of nanowire (see Fig.1b) [13]. When we fabricate a d.c. SQUID device using TI nanowire and conventional superconducting electrodes, the $I_c$ modulation with perpendicular magnetic field would reflect the current-phase relation as a function of gate voltage and temperature.

Thirdly, the existence of Majorana fermion states occurring at two ends of TI NW [22] can be directly probed by measuring zero-bias conductance peak (ZBCP) structure in the superconducting proximity region of TI nanowire (see Fig.1c). When a tunneling-type junction is formed on TI nanowire, which is in highly transparent contact with a superconducting electrode, the differential conductance, $dI/dV$, measurement would reveal ZBCP due to a finite density of state in the normal-metal/insulator/TI nanowire/superconductor junction. Since the one-dimensional subband gap can be induced in the nanowire with applying an external magnetic field, $B_{axial}$, along the TI nanowire axis, a periodic modulation of the ZBCP with a periodicity of $\Phi_0$ is expected. Thus the ZBCP as a function of gate voltage ($V_g$) and $B_{axial}$ will verify the existence of Majorana fermion in TI nanowire.

## 4. DISCUSSIONS

**High quality of TI nanowire :** In our experimental design for probing Majorana states in TI nanowire, it is essential to resolve the helical surface state residing on the surface of TI nanowire using electrical transport measurements. This would be plausible only when the Fermi energy, $E_F$, is located near Dirac point. Thus it is required that TI nanowires should exhibit an ambipolar gate dependence to clearly show its Dirac point and high mobility enough to exhibit Aharonov-Bohm oscillations with the axial magnetic field. Previous studies indicate that several TI nanowires, such as $Bi_2Se_3$ [20], $Bi_2Te_3$ [24], $Ag_2Se$ [25], $Bi_{1.5}Sb_{0.5}Te_{1.7}Se_{1.3}$ [29], and $(Bi_{1-x}Sb_x)_2Se_3$, can be a promising candidate to satisfy this requirement condition.

**Large critical field of superconducting electrode :** We aim to combine the helical surface state of TI nanowire with the superconducting proximity effect. Especially, when the magnetic flux through the nanowire section is given by $\Phi_0/2 = h/2e$, we can resolve the surface state in TI nanowires. Usually the magnetic field corresponding to $\Phi = h/2e$ ranges from 0.1 to 0.2 Tesla, which was already known from our previous work [25]. Since the superconductivity of Al is easily broken in this field range, it is required to use another superconducting material with higher critical magnetic field, $H_C$, than 0.1 Tesla. This requirement also rules out Pb and PbIn alloy as a candidate material for superconductor [28]. Thus it would be more promising to try Nb, NbN, Ta, TaN as a superconductor. Though these materials exhibits very high $H_C$ (> 1 Tesla), the observation of supercurrent through TI nanowires would become another issue.

**Highly transparent contact between TI and superconductor :** Formation of highly transparent superconducting contact is quite critical to observe the supercurrent in nano-hybrid junctions. Since the nanostructures are made of semiconductors, it is very hard to avoid a formation of Schottky barrier at the interface between semiconductor nanostructure and superconducting electrode [27]. This issue will be revisited with TI nanowires. Though it is well known that Schottky barrier is dependent of the work function of metallic electrode, carrier concentration, and band gap etc., the experimental results reveals that it is quite empirical (case-by-case) to get a good fabrication recipe for the highly transparent superconducting contact. So far we have obtained successful recipes using Al [26] and PbIn [28] superconductors, but othere recipes with Nb, NbN, Ta, and TaN, are urgently needed. Our experimental result using $(Bi_{1-x}Sb_x)_2Se_3$ TI nanowire is shown in Fig. 2, where $I_c$ reaches up to ~ 1 μA at $T$ = 300 mK. Here, the $I_c R_n$ product is obtained to be ~ 23 μV, in which $R_n$ is normal-state resistance of the JJ.

In summary, detecting Majorana fermions in topological insulator nanowires and its application to topological supercurrent qubit would be a quite exciting scientific achievement in itself. Moreover, this research work would pave a novel way to a gate-tunable, scalable, and topologically-protectable qubit.

# ACKNOWLEDGMENT

This work was supported by the National Research Foundation of Korea through the Basic Science Research Program (Grant 2018R1A3B1052827).

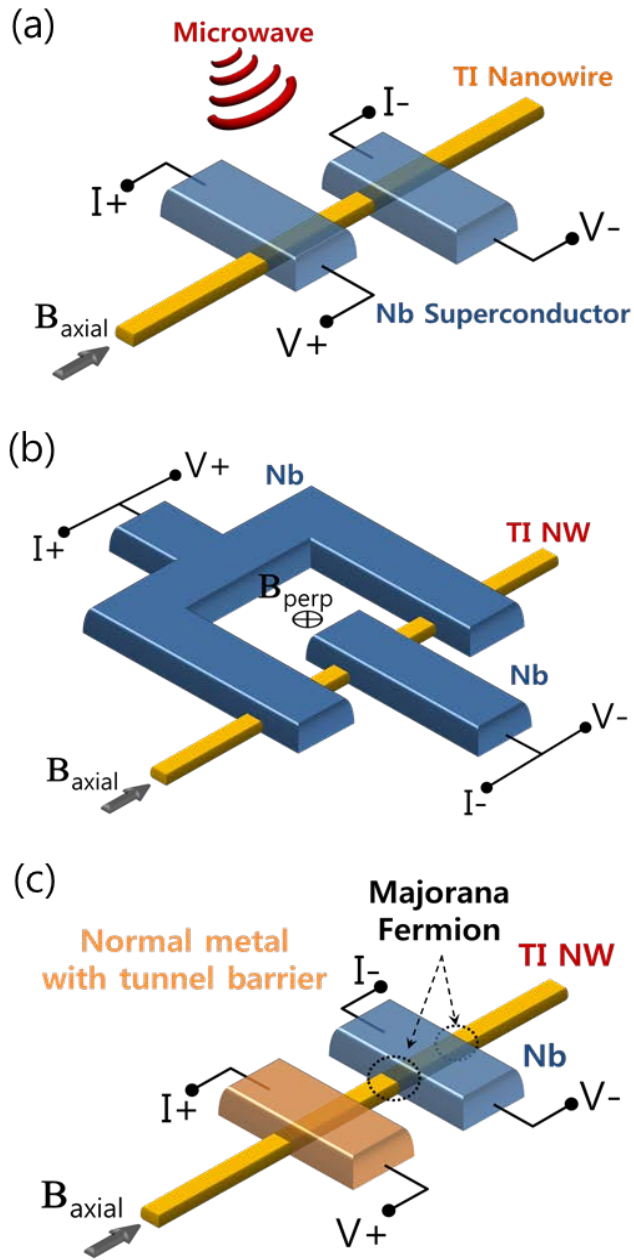

Fig. 1. Schematic view of our experimental designs for the measurements of (a) anomalous Shapiro steps, (b) current-phase relation, and (c) zero-bias conductance peak as an evidence of Majorana fermion state in topological insulator nanowire.

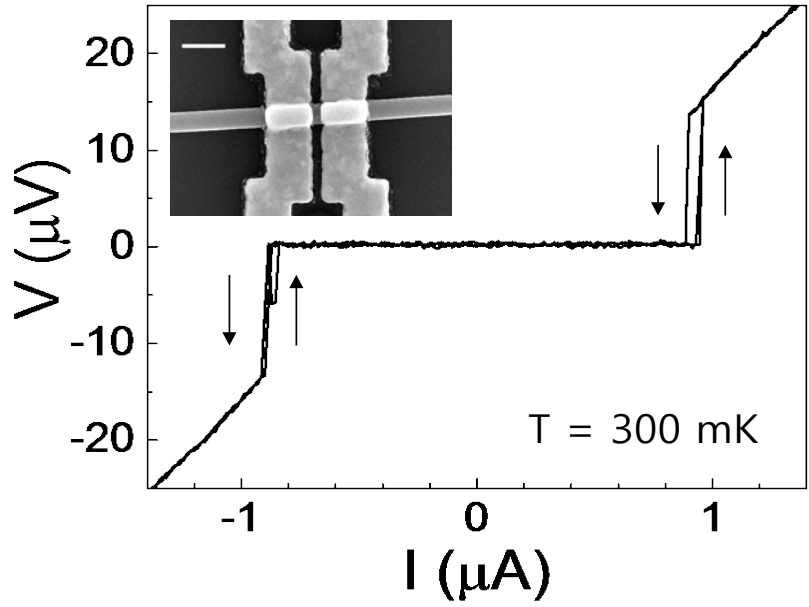

Fig. 2. Current vs. voltage characteristic curve obtained from $(Bi_{1-x}Sb_x)_2Se_3$ TI nanowire contacted with PbIn superconducting electrodes. The arrows indicate a hysteretic behavior. Inset: representative scanning electron microscopy image of TI nanowire-based JJ. Scale bar indicates 1 μm.